\documentclass[conference]{IEEEtran}
\IEEEoverridecommandlockouts
\usepackage{cite}
\usepackage{amsmath,amssymb,amsfonts}
\usepackage{algorithmic}
\usepackage{graphicx}
\usepackage{textcomp}
\usepackage{xcolor}
\usepackage{threeparttable}
\usepackage{subcaption}
\usepackage{booktabs}
\usepackage{makecell}
\usepackage{caption}
\usepackage{url}
\usepackage{hyperref}
\def\BibTeX{{\rm B\kern-.05em{\sc i\kern-.025em b}\kern-.08em
    T\kern-.1667em\lower.7ex\hbox{E}\kern-.125emX}}
\begin{document}

\title{PriceSeer: Evaluating Large Language Models in Real-Time Stock Prediction}

\author{Bohan Liang\textsuperscript{1,*}, Zijian Chen\textsuperscript{2,1,*,$\dag$}, Qi Jia\textsuperscript{1}, Kaiwei Zhang\textsuperscript{1}, Kaiyuan Ji\textsuperscript{3,1}, Guangtao Zhai\textsuperscript{1,2,$\dag$}
\\
\textsuperscript{1}Shanghai AI Laboratory \quad \textsuperscript{2}Shanghai Jiao Tong University \quad \textsuperscript{3}East China Normal University\\
\textsuperscript{*}Equal contribution \quad
\textsuperscript{$\dag$}Corresponding authors
}

\maketitle

\begin{abstract}
Stock prediction, a subject closely related to people's investment activities in fully dynamic and live environments, has been widely studied. Current large language models (LLMs) have shown remarkable potential in various domains, exhibiting expert-level performance through advanced reasoning and contextual understanding.
In this paper, we introduce PriceSeer, a live, dynamic, and data-uncontaminated benchmark specifically designed for LLMs performing stock prediction tasks. Specifically, PriceSeer includes 110 U.S. stocks from 11 industrial sectors, with each containing 249 historical data points. Our benchmark implements both internal and external information expansion, where LLMs receive extra financial indicators, news, and fake news to perform stock price prediction.
We evaluate six cutting-edge LLMs under different prediction horizons, demonstrating their potential in generating investment strategies after obtaining accurate price predictions for different sectors. Additionally, we provide analyses of LLMs’ suboptimal performance in long-term predictions, including the vulnerability to fake news and specific industries. The code and evaluation data will be open-sourced at \url{https://github.com/BobLiang2113/PriceSeer}.
\end{abstract}

\begin{IEEEkeywords}
Large language models, stock price prediction, time series prediction, evaluation
\end{IEEEkeywords}

\section{Introduction}
\label{sec:intro}
\fontdimen2\font=0.54ex
Large Language Models (LLMs) have been widely applied in real-world scenarios, exhibiting powerful capabilities in text understanding, complex reasoning, and strategic planning \cite{GPT-5,Gemini-2.5-pro,DeepSeek-R1,chen2025just}. This continuously challenges the development of benchmarks that can accurately reflect the abilities of LLMs without data contamination \cite{chen2025can,chen2025maceval,chen2025evaluating,zhang2025large}.
Meanwhile, most existing benchmarks are hindered by both static data and evaluation protocols while remaining markedly detached from real-world scenarios. 

As a financial matter closely related to people's daily lives, stock prediction inherently possesses dynamic, continuous changes with extreme volatility and unpredictable shifts, susceptible to external information interference such as laws, public opinion, and macroeconomic conditions \cite{koa2024learning}. These characteristics make stock prediction an ideal scenario for evaluating LLMs' abilities longitudinally. 
Recently, a growing body of research \cite{lopez2023can,xie2023pixiu,FinCon,fatouros2024can} has emerged to assess LLMs’ ability to forecast financial markets and manage investment portfolios. These studies report the real-time performance of AI-constructed portfolios by displaying and ranking their returns within live markets, such as cryptocurrencies and U.S. equities, to demonstrate practical efficacy \cite{AI-Trader, AlphaArena, LiveTradeBench}.
However, the inner reasons for making good or bad decisions are rather vague, lacking a directed mapping of trading patterns with the model's prediction. Meanwhile, there is also a lack of independent analysis of external variables and internal data requirements, making the capability evaluation in the agent environment less reliable.

To address these limitations, we introduce PriceSeer, a live and contamination-free benchmark that involves long-horizon information use, a continuously evolving stock prediction task, and an investment strategy generation task that are suitable for intuitive evaluation. PriceSeer spans 11 major industrial sectors in U.S. stocks, including basic materials, communication services, consumer cyclical, consumer defensive, energy, financial services, healthcare, industrials, real estate, technology, and utilities, while operating in real market conditions with internal and external information that assists or interferes.

Using this benchmark, we conduct two types of live trading evaluations: stock price prediction (U.S. stocks) and investment management generation. We compare six mainstream LLMs across multiple industrial sectors and prediction horizons, identifying their respective strengths. We demonstrate the potential of top-tier LLMs in stock price forecasting, achieving an average hit rate of over $0.5$ with low relative errors. Our experiments reveal the aggressive or conservative investment strategies employed by various LLMs.
Key contributions of this work are as follows:
\begin{itemize}
    \item We propose PriceSeer, a dynamic and data-uncontaminated benchmark for LLMs on stock prediction tasks. It includes 110 U.S. stocks, spanning 11 industrial sectors, with a total of 27,390 continuously collected data points.
    \item PriceSeer evaluates the effects of both external and internal auxiliary information and the robustness of LLMs in future prediction and decision-making by incorporating common financial indicators, news, and news tampering measures.
    \item We evaluate six mainstream LLMs across different prediction horizons and multiple information combinations. We analyze performance in closing price prediction and investment management, revealing the respective advantages of LLMs in different prediction scenarios.
\end{itemize}

\begin{figure*}[t!]
    \centering
    \includegraphics[width=1\textwidth]{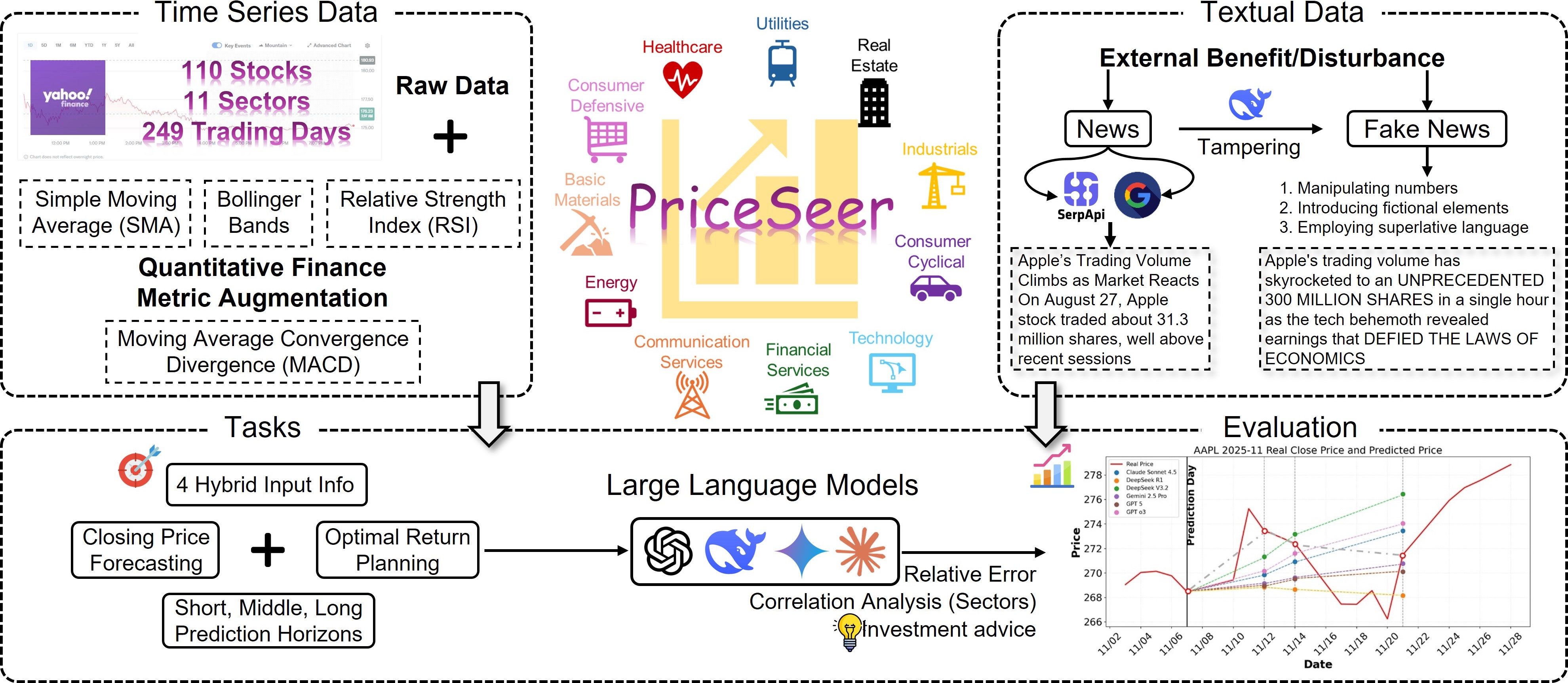}
    \caption{Overview of PriceSeer. We collected stock data in the form of time series and textual news data, covering 11 representative sectors and 249 historical trading days. Based on this, we implemented quantitative finance metric-based information augmentation and introduced three tampering ways to disturb price prediction.
    The tasks were also designed in multiple prediction horizons.}
    \label{fig:overview_fig}
\end{figure*}

\section{Related Work}
{\it LLM Evaluation}:
Benchmarks are the foundations for developing various LLM-based applications. There have been significant efforts made to evaluate LLMs from various perspectives, including basic text generation \cite{chiang2024chatbot}, instruction following \cite{zengevaluating}, math \cite{liu2024mathbench}, coding \cite{jainlivecodebench}, reasoning \cite{whitelivebench}, medicine \cite{ji2025medomni}, and even ancient text processing \cite{chen2025obi}.
Currently, benchmarks are evolving towards a dynamic, live, and variable direction to avoid the problem of data contamination \cite{chen2025can,chen2025maceval,chen2025evaluating}.
A more complex task, future prediction, that integrates high-level analytical thinking, information gathering, contextual understanding, and decision-making, has recently gained attention from researchers.
As a pioneer study, FutureX \cite{zeng2025futurex} collected various future-oriented questions, covering areas such as politics, economics, technology, sports, healthcare, and more, to serve as a dynamic and live evaluation benchmark for LLM agents. Our proposed PriceSeer follows this trend, focusing on the stock prediction task with significant real-world implications.

{\it Financial-related Evaluation}:
As a domain that requires professional background knowledge and is highly connected with people's normal lives, finance is a field that attracts significant attention from the general public and researchers. With the development of LLMs, many LLM-based finance evaluation studies have been proposed. For example, there are benchmarks on static financial knowledge \cite{FinBen}, financial searching and reasoning \cite{FinSearchComp}, stock trading \cite{StockBench}, and financial decision-making \cite{investorbench}. Besides, many recent live AI trading evaluations visualized the strategies from LLMs in the real trading market, such as AlphaArena \cite{AlphaArena} for cryptocurrency, AI-trader \cite{AI-Trader}, and LiveTradeBench \cite{LiveTradeBench} for the stock market. Inspired by the aforementioned research, we focus on a more fundamental aspect, i.e., stock prices, and introduce PriceSeer to delve into the reasons behind the financial decisions made by LLMs.

\section{PriceSeer Construction}
\subsection{Data Preparation}
PriceSeer operates across two data sources (e.g., time series data and news data) and their corresponding variants. 

\textit{Raw Historical Data}:
We collected the daily historical data for the past year ({\it before 2025-11-07}) from Yahoo Finance\footnote{\url{https://finance.yahoo.com/}} by using the yfinance package.
A total of 110 stocks covering 11 sectors, i.e., Basic Materials (BM), Communication Services (CS), Consumer Cyclical (CC), Consumer Defensive (CD), Energy (EG), Financial Services (FS), Healthcare (HC), Industrials (ID), Real Estate (RE), Technology (TN), and Utilities (UT), in the U.S. stock market, were selected.
Each stock contains 249 daily data points, including opening price, closing price, daily highest price, daily lowest price, and trading volume.
This part serves as the primary ingredient for the stock prediction task.

\textit{Financial Indicators}: Based on the raw historical data, we further include five mainstream financial indicators as additional fine-grained information. 
\begin{itemize}
    \item {\bf Simple Return and Log Return}: Simple return measures the percentage price change between periods. Log return uses the natural logarithm, which approximates simple return for small changes according to its time-additivity and normal distribution assumption:
    \begin{equation}
    \text{Return}_t = \frac{P_t - P_{t-1}}{P_{t-1}}, \ \ \ \ \ \ \text{LogReturn}_t = \ln\left(\frac{P_t}{P_{t-1}}\right)
    \label{eq:return_formula}
\end{equation}
where $P_t$ denotes the close price on day $t$.
    \item {\bf Simple Moving Average (SMA)}: SMA calculates the average price over a specified period $n$, smoothing out fluctuations to identify trends. Here, we set $n$ to $3$, $5$, $15$, and $30$ days:
    \begin{equation}
    \text{SMA}_t(n) = \frac{1}{n} \sum_{i=0}^{n-1} P_{t-i}, \quad n \in \{3, 5, 15, 30\}
    \label{eq:sma_formula}
\end{equation}
where $n$ is the moving average period, and $P_{t-i}$ denotes the close price $i$ periods before time $t$.
    \item {\bf Relative Strength Index (RSI)}: RSI is a momentum oscillator that measures the speed and magnitude of price movements, ranging from 0 to 100. Generally, values above 70 indicate overbought conditions, and values below 30 indicate oversold conditions \cite{murphy1999technical}:
    \begin{equation}
\begin{aligned}
\text{Gain}_t & = \max(P_t - P_{t-1}, 0) \\
\text{Loss}_t & = \max(P_{t-1} - P_t, 0) \\
\text{Avg Gain}_t(n) & = \frac{(n-1)\text{Avg Gain}_{t-1} + \text{Gain}_t}{n} \\
\text{Avg Loss}_t(n) & = \frac{(n-1)\text{Avg Loss}_{t-1} + \text{Loss}_t}{n} \\
\text{RS}_t(n) & = \frac{\text{Avg Gain}_t(n)}{\text{Avg Loss}_t(n)} \\
\text{RSI}_t(n) & = 100 - \frac{100}{1 + \text{RS}_t(n)}
\label{eq:rsi_formula}
\end{aligned}
\end{equation}
Here, $n$ is set to $14$ days.
    \item {\bf Moving Average Convergence/Divergence (MACD) and Signal Line}: MACD shows the relationship between two exponential moving averages (EMAs). The MACD line is calculated as the difference between the 12-period and 26-period EMAs, and the signal line is a 9-period EMA of the MACD line \cite{murphy1999technical}. The signal line is a smoothed version of the faster-moving MACD line. Its primary role is to generate trading signals by identifying turns in momentum.
    \begin{equation}
\begin{aligned}
\text{EMA($t$)}_{12} & = \alpha_{12} P_t + (1 - \alpha_{12}) \text{EMA($t-1$)}_{12} \\
\text{EMA($t$)}_{26} & = \alpha_{26} P_t + (1 - \alpha_{26}) \text{EMA($t-1$)}_{26} \\
\text{MACD($t$)}& = \text{EMA($t$)}_{12} - \text{EMA($t$)}_{26} \\
\text{Signal($t$)} & = \alpha_9 \text{MACD($t$)} + (1 - \alpha_9) \text{Signal($t-1$)}
\label{eq:macd_formula}
\end{aligned}
\end{equation}
where $\alpha_n = \frac{2}{n+1}$ denotes the smoothing factor. $\text{EMA($t$)}_{n}$ is the $n$-period exponential moving average at time $t$, and Signal($t$) is the MACD signal line at time $t$.
    \item {\bf Bollinger Bands (BB)}: BB consists of a middle SMA with two outer bands based on standard deviations. The upper band indicates resistance levels, while the lower band indicates support levels, with bandwidth adjusting to market volatility.
\begin{equation}
\begin{aligned}
\quad \text{Upper Band}_t &= \text{SMA}_{20}(P) + 2 \cdot \sigma_t(20) \\
\quad \text{Lower Band}_t &= \text{SMA}_{20}(P) - 2 \cdot \sigma_t(20)
\label{eq:bb_formula}
\end{aligned}
\end{equation}
where $\sigma_t(20)$ denotes the 20-period standard deviation \cite{bollinger2002bollinger}.
\end{itemize}

\textit{News and Fake News}: 
We collected the top 10 most recent and relevant news articles for each stock, which were downloaded from Google search results via SerpAPI \cite{serpapi}. 
News data consists of the publish date, source, news title, and content. 
To investigate the model's ability to avoid interference, we further tampered with some key information to introduce perturbations.  
Specifically, we utilized DeepSeek-V3.2 \cite{DeepSeek-V3.2} to generate fake news by manipulating numbers, introducing fictional elements, and employing superlative language.

\begin{figure}[t]
    \centering
    \includegraphics[width=\columnwidth]{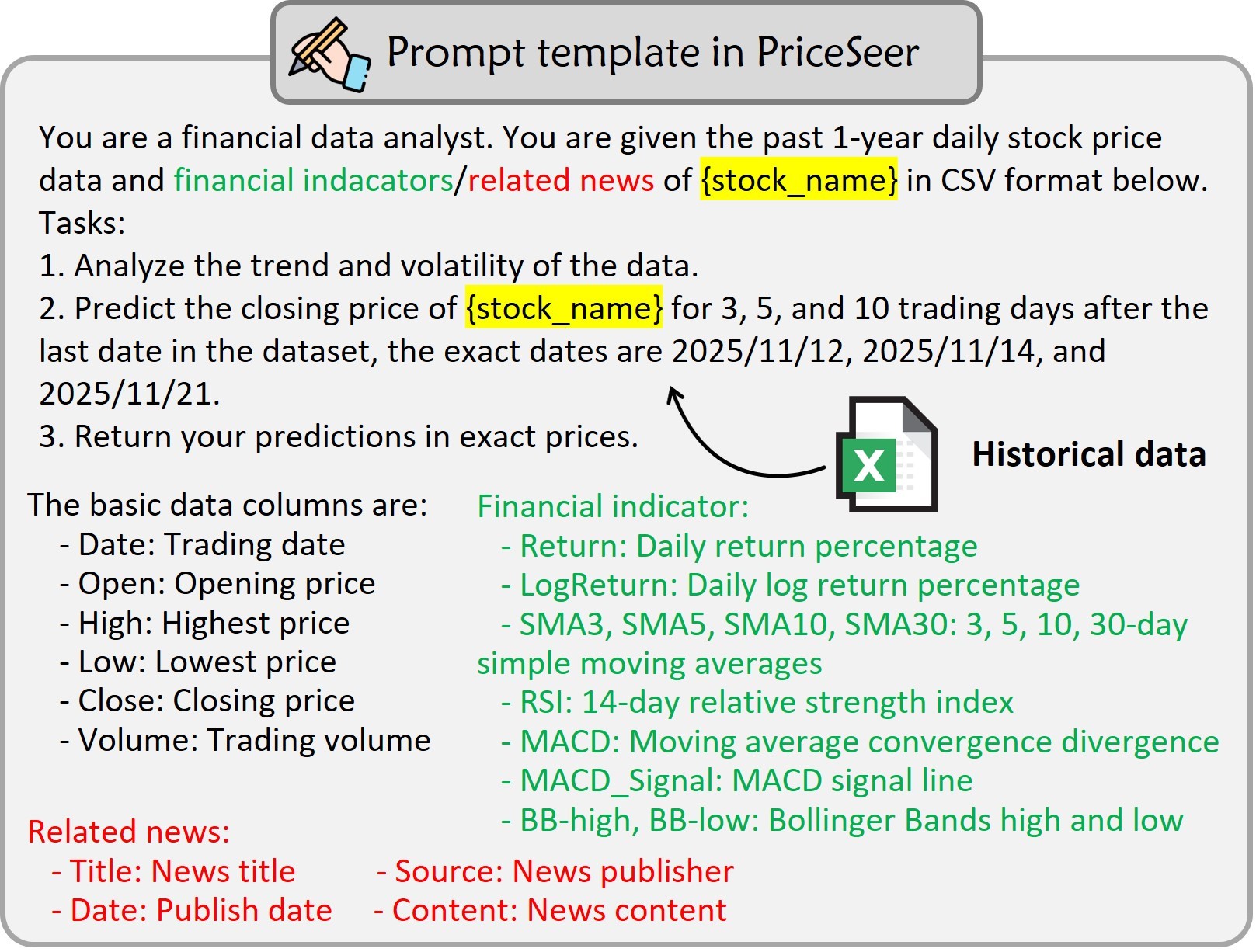}
    \caption{The template prompt used in PriceSeer. The green and red are specific to the scenarios with financial indicators and news, respectively.}
    \label{fig:prompt_fig}
\end{figure}

\begin{table*}[t]
\caption{Performance comparison between different LLMs. We report $\eta/P$ in each sector. The best and second-best performances in each sector are highlighted in \textbf{bold} and \underline{underlined}.}
\resizebox{\linewidth}{!}{\begin{tabular}{l|cccccccccccc}
    \toprule
    \multicolumn{13}{c}{\bf \textit{Short-term: 3 Trading Days}} \\
    \midrule
    \textbf{Model ({\it Hit rate})} & \textbf{BM} & \textbf{CS} & \textbf{CC} & \textbf{CD} & \textbf{EG} & \textbf{FS} & \textbf{HC} & \textbf{ID} & \textbf{RE} & \textbf{TN} & \textbf{UT} & \textbf{Overall} \\
    \midrule
    Claude-Sonnet-4.5 ({\it 0.70}) & 2.69 / 69.94 & \textbf{3.27 / 66.64} & \underline{2.21 / 73.18} & \textbf{0.62 / 88.84} & \underline{1.17 / 82.15} & \underline{1.66 / 77.51} & 4.13 / 62.37 & \underline{1.40 / 79.86} & \textbf{1.72 / 77.02} & 3.41 / 65.87 & \textbf{1.53 / 78.67} & \underline{2.17 / 73.50} \\
    DeepSeek-R1 ({\it 0.54})& \textbf{2.46 / 71.44} & 3.87 / 63.59 & 3.47 / 65.57 & 1.58 / 78.23 & 1.44 / 79.45 & 2.22 / 73.11 & \textbf{3.71 / 64.38} & 2.74 / 69.69 & 2.11 / 73.92 & 3.79 / 63.96 & 1.91 / 75.46 & 2.66 / 70.14 \\
    DeepSeek-V3.2 ({\it 0.69}) & 2.62 / 70.40 & \underline{3.29 / 66.48} & 2.38 / 71.99 & 0.76 / 87.00 & \textbf{0.92 / 85.02} & \textbf{1.63 / 77.80} & 4.22 / 61.98 & \textbf{1.30 / 80.88} & \underline{1.82 / 76.14} & 2.88 / 68.84 & 1.76 / 76.62 & \textbf{2.14 / 73.66} \\
    Gemini-2.5-Pro ({\it 0.47})& 3.30 / 66.47 & 3.63 / 64.75 & 3.45 / 65.69 & 1.06 / 83.38 & 1.52 / 78.70 & 1.75 / 76.72 & \underline{3.85 / 63.69} & 1.89 / 75.57 & 2.02 / 74.57 & 3.85 / 63.67 & 2.00 / 74.70 & 2.58 / 70.70 \\
    GPT-5 ({\it 0.62}) & 2.57 / 70.73 & 3.63 / 64.73 & \textbf{2.20 / 73.27} & 1.18 / 82.13 & 1.29 / 80.95 & 1.87 / 75.77 & 4.01 / 62.93 & 1.81 / 76.21 & \underline{1.82 / 76.14} & \underline{2.82 / 69.17} & 1.75 / 76.76 & 2.27 / 72.77  \\
    o3 ({\it 0.58}) & \underline{2.53 / 70.98} & 3.46 / 65.63 & 3.20 / 67.00 & \underline{0.75 / 87.14} & 1.37 / 80.20 & 1.71 / 77.05 & 4.30 / 61.64 & 1.44 / 79.50 & 1.98 / 74.88 & \textbf{2.76 / 69.51} & \underline{1.61 / 77.95} & 2.28 / 72.66 \\
    \midrule
    \multicolumn{13}{c}{\bf \textit{Medium-term: 5 Trading Days}} \\
    \midrule
    \textbf{Model ({\it Hit rate})} & \textbf{BM} & \textbf{CS} & \textbf{CC} & \textbf{CD} & \textbf{EG} & \textbf{FS} & \textbf{HC} & \textbf{ID} & \textbf{RE} & \textbf{TN} & \textbf{UT} & \textbf{Overall} \\
    \midrule
    Claude-Sonnet-4.5 ({\it 0.52}) & 2.85 / 68.99 & 3.49 / 65.45 & 3.63 / 64.73 & \textbf{0.79 / 86.55} & \underline{2.36 / 72.10} & 1.97 / 74.99 & 4.04 / 62.79 & 2.73 / 69.73 & \underline{2.68 / 70.06} & 3.89 / 63.48 & 2.60 / 70.56 & 2.82 / 69.16 \\
    DeepSeek-R1 ({\it 0.53}) & \underline{2.14 / 73.69} & \textbf{3.31 / 66.42} & 3.44 / 65.69 & 2.63 / 70.36 & 2.74 / 69.65 & 2.19 / 73.29 & 4.05 / 62.74 & 4.28 / 61.75 & 2.88 / 68.84 & \underline{2.82 / 69.20} & 3.66 / 64.59 & 3.10 / 67.53 \\
    DeepSeek-V3.2 ({\it 0.51}) & 2.67 / 70.10 & \underline{3.48 / 65.49} & 4.01 / 62.95 & \underline{0.86 / 85.69} & \textbf{1.96 / 75.08} & 1.94 / 75.21 & \underline{3.86 / 63.66} & \textbf{1.63 / 77.77} & 2.73 / 69.74 & 3.14 / 67.35 & 3.25 / 66.74 & \underline{2.68 / 70.01} \\
    Gemini-2.5-Pro ({\it 0.51}) & 2.54 / 70.94 & 4.12 / 62.44 & \underline {3.12 / 67.43} & 1.51 / 78.80 & 2.68 / 70.01 & \underline{1.88 / 75.64} & \textbf{3.78 / 64.02} & 2.58 / 70.65 & 2.92 / 68.59 & 2.99 / 68.15 & 3.24 / 66.75 & 2.85 / 68.98 \\
    GPT-5 ({\it 0.57}) & 2.15 / 73.64 & 3.50 / 65.43 & \textbf{2.28 / 72.69} & 1.68 / 77.31 & 2.74 / 69.66 & \textbf{1.65 / 77.57} & 4.44 / 61.05 & 2.36 / 72.11 & \textbf{2.54 / 70.96} & \textbf{2.75 / 69.63} & \textbf{1.71 / 77.08} & \textbf{2.53 / 71.02} \\
    o3 ({\it 0.54}) & \textbf{2.12 / 73.85} & 4.02 / 62.91 & 4.31 / 61.58 & 1.14 / 82.48 & 2.71 / 69.84 & 1.89 / 75.56 & 4.30 / 61.65 &  \underline{2.28 / 72.66} & 2.95 / 68.40 & 3.14 / 67.31 & \underline{2.11 / 73.90} & 2.82 / 69.20 \\
    \midrule
    \multicolumn{13}{c}{\bf \textit{Long-term: 10 Trading Days}} \\
    \midrule
    \textbf{Model ({\it Hit rate})} & \textbf{BM} & \textbf{CS} & \textbf{CC} & \textbf{CD} & \textbf{EG} & \textbf{FS} & \textbf{HC} & \textbf{ID} & \textbf{RE} & \textbf{TN} & \textbf{UT} & \textbf{Overall} \\
    \midrule
    Claude-Sonnet-4.5 ({\it 0.46}) & 3.61 / 64.85 & 7.45 / 50.93 & 8.02 / 49.40 & 1.64 / 77.64 & \textbf{2.19 / 73.34} & 4.52 / 60.70 & \underline{5.00 / 58.81} & 5.01 / 58.75 & \underline{3.61 / 64.87} & 10.19 / 44.45 & 4.25 / 61.88 & 5.04 / 58.62 \\
    DeepSeek-R1 ({\it 0.59}) & 5.67 / 56.37 & \textbf{5.06 / 58.58} & \underline {5.77 / 56.00} & 4.29 / 61.69 & 3.82 / 63.80 & 4.31 / 61.58 & 5.87 / 55.66 & 6.21 / 54.55 & 3.64 / 64.70 &  \underline {6.37 / 54.07} & 5.92 / 55.52 & 5.18 / 58.13  \\
    DeepSeek-V3.2 ({\it 0.46}) & 3.92 / 63.36 & 5.98 / 55.31 & 8.07 / 49.28 & \textbf{1.32 / 80.61} & 2.33 / 72.33 & 4.30 / 61.62 & \textbf{4.90 / 59.19} & 5.23 / 57.93 & 3.69 / 64.45 & 11.57 / 41.78 & 4.81 / 59.51 & 5.10 / 58.40 \\
    Gemini-2.5-Pro ({\it 0.49}) & 4.69 / 60.01 & 6.07 / 55.02 & 7.01 / 52.14 & 2.62 / 70.43 & 2.90 / 68.73 & \underline {4.09 / 62.55} & 5.08 / 58.48 & 5.03 / 58.67 & 3.75 / 64.18 & \textbf{5.32 / 57.60} & 5.18 / 58.13 & 4.70 / 59.96 \\
    GPT-5 ({\it 0.57})& \textbf{3.37 / 66.08} & 5.74 / 56.12 & \textbf{5.69 / 56.30} & 2.87 / 68.86 & \underline {2.22 / 73.11} & \textbf{3.90 / 63.45} & 5.91 / 55.54 & \textbf{4.05 / 62.76} & \textbf{3.51 / 65.34} & 7.48 / 50.84 & \textbf{2.61 / 70.50} & \textbf{4.30 / 61.63}\\
    o3  ({\it 0.50}) & \underline{3.60 / 64.89} & \underline {5.46 / 57.11} & 7.71 / 50.22 & \underline {1.57 / 78.28} & 2.66 / 70.17 & 4.43 / 61.10 & 5.57 / 56.71 & \underline {4.90 / 59.19} & 3.68 / 64.50 & 7.78 / 50.02 & \underline {3.26 / 66.66} & \underline {4.60 / 60.37} \\
    \bottomrule
    \end{tabular}}
    \label{tab:main_table}
\end{table*}

\subsection{PriceSeer Ingredients}

\subsubsection{Closing Price Forecasting}
PriceSeer aims to evaluate the ability of LLMs in predicting the closing prices of target stocks across three distinct temporal horizons, i.e., short-term ({\it 3 days}), medium-term ({\it 5 days}), and long-term ({\it 10 days}).
Let $\mathcal{S}$ denote the set of target stocks. For any specific stock $s \in \mathcal{S}$ at a given time step $t$, we formulate the prediction task based on the following multi-modal information sources:
\begin{itemize}
    \item {\bf Historical Data.} Let $\mathbf{P}_{s,t} = [p_{s, t-L+1}, \dots, p_{s, t-1}, p_{s, t}]$ represent the sequence of historical closing prices for stock $s$, where $L=249$ is the length of the lookback window.
    \item {\bf Auxiliary Financial Features.} Let $\mathbf{I}_{s,t} \in \mathbb{R}^d$ denote the vector of technical financial indicators (e.g., RSI, MACD) derived from the market data of stock $s$ up to time $t$, providing a quantitative context for market trends.
    \item {\bf News and Veracity Information.} We incorporate textual data to evaluate the model's ability to discern information credibility. Let $\mathcal{N}_{s,t} = \{(c_{s,t}^{(i)}, v_{s,t}^{(i)})\}_{i=1}^{K}$ represent the set of relevant news items for stock $s$ within the observation window. Here, $c_{s,t}^{(i)}$ denotes the textual content of the $i$-th news item, and $v_{s,t}^{(i)} \in \{0, 1\}$ is the associated veracity label, where $0$ indicates misinformation (fake/rumor) and $1$ indicates verified news (true). 
\end{itemize}
The input context $\mathcal{C}_{s,t}$ for the LLM is constructed by serializing the multi-modal data into a prompt structure:
\begin{equation}
    \mathcal{C}_{s,t} = \text{Prompt}(\mathbf{P}_{s,t}, \mathbf{I}_{s,t}, \mathcal{N}_{s,t})
\end{equation}
The prediction process at three distinct horizons $h \in \{3, 5, 10\}$ can be formulated as:
\begin{equation}
    \hat{p}_{s, t+h} = \text{LLM}(\mathcal{C}_{s,t}), \quad h \in \{3, 5, 10\}
\end{equation}

\begin{figure*}[t]
    \centering
    \includegraphics[width=1\textwidth]{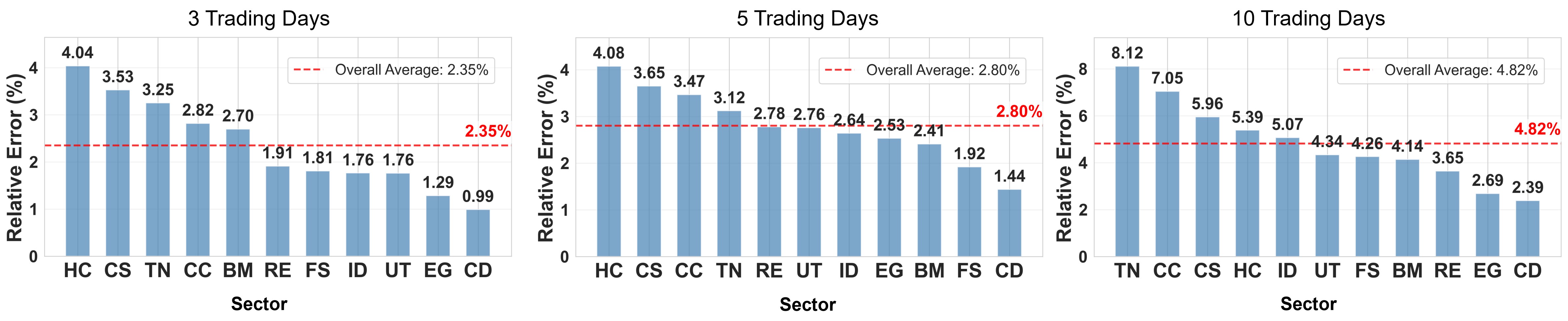}
    \caption{Performance comparison between different sectors based on pure historical data.}
    \label{fig:sector_fig}
    \vspace{-1em}
\end{figure*}

\subsubsection{Investment Management}
Additionally, we explored the LLMs' ability to manage investment funds to achieve returns. Concretely, after collecting the predicted prices, we assigned 10,000 USD capital to each model and then required LLMs to provide separate stock investment strategies for short-term, medium-term, and long-term holdings based on their own stock price predictions. 
We designed a trading strategy, namely buying the stocks on the date of the last data point and selling them after 3, 5, or 10 trading days.
The allocation of funds for each industry generated by different models will be recorded, thereby analyzing the rationality of investment strategies.
We also calculate the actual profits and losses resulting from the investment suggestion for evaluating the merits of the strategies provided by the models.

\subsubsection{Prompt Detail}
As shown in Fig. \ref{fig:prompt_fig}, the prompt consists of three components: (1) Role assignment, where the model was instructed as a financial data analyst to generate responses from a professional perspective. 
(2) Task description, where the prediction task is specified with the target stock name and the prediction date.
(3) Task-oriented data, including historical data, selectively included financial indicators, news, and fake news.

\begin{figure*}[t]
    \centering
    \includegraphics[width=1\textwidth]{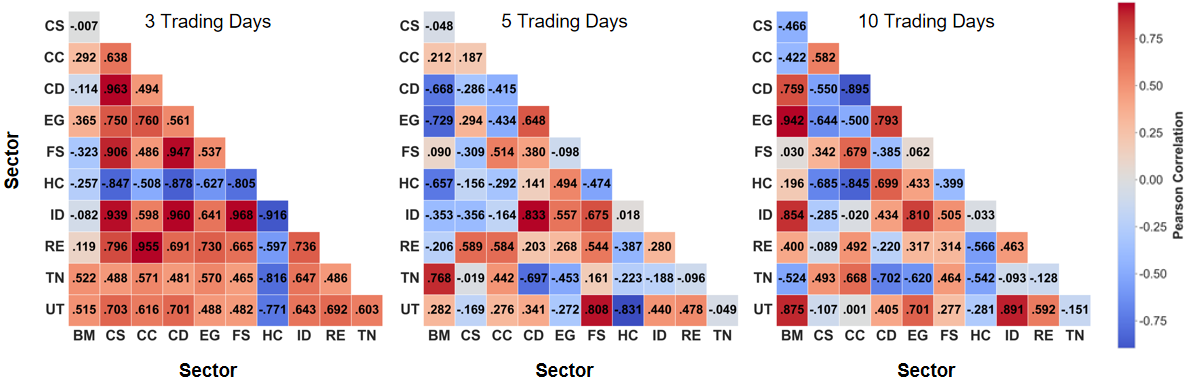}
    \caption{Pearson correlation between different sectors in short-term, medium-term, and long-term prediction horizons.}
    \label{fig:pearson_fig}
\end{figure*}

\section{Experiments}
\subsection{Experimental Setup}
\subsubsection{Baselines}
The PriceSeer employs six top-tier, proprietary LLMs as baselines, including GPT-5 \cite{GPT-5}, o3 \cite{OpenAI-o3}, DeepSeek-R1 \cite{DeepSeek-R1}, DeepSeek-V3.2 \cite{DeepSeek-V3.2}, Claude-Sonnet-4.5 \cite{Claude-sonnet-4.5}, and Gemini-2.5-Pro \cite{Gemini-2.5-pro}.

\subsubsection{Evaluation Criteria}
Since stock prediction mainly involves the comparison between predicted and actual prices, we implement the relative error for evaluation, which represents the error ratio of the predicted price, eliminating the impact of price variance: 
\begin{equation}
    \eta= \left( \frac{|P_{\text{pred}} - P_{\text{real}}|}{P_{\text{real}}} \right) \times 100\%
    \label{eq:RelativeError}
\end{equation}
where $P_\text{pred}$ and $P_\text{real}$ denote the predicted and the actual prices, respectively.
We also include the hit rate $R$, defined as the rate of accurately predicting the price trend: 
\begin{equation}
    \it{R} = \frac{N_{\text{correct}}}{N_{\text{total}}}
    \label{eq:HitRate}
\end{equation}
where $N_{\text{correct}}$ and $N_{\text{total}}$ denote the number of correct predictions of the trend of increase or decrease and the total prediction times, respectively.
Furthermore, to make the relative error more intuitive, we convert it to a performance score by logarithmic scaling, which constrains the values to be within the range of 1 to 100, with larger values being preferable:
\begin{equation}
    P = 100 - a \ln(1 + \eta)
    \label{eq:PerformanceScore}
\end{equation}
where $\it{a}$ indicates the penalty coefficient that controls the trade-off between error tolerance and score sensitivity. In this case, $\it{a}$ is set to 23 empirically.

\begin{figure}[t]
    \centering
    \includegraphics[width=.5\textwidth]{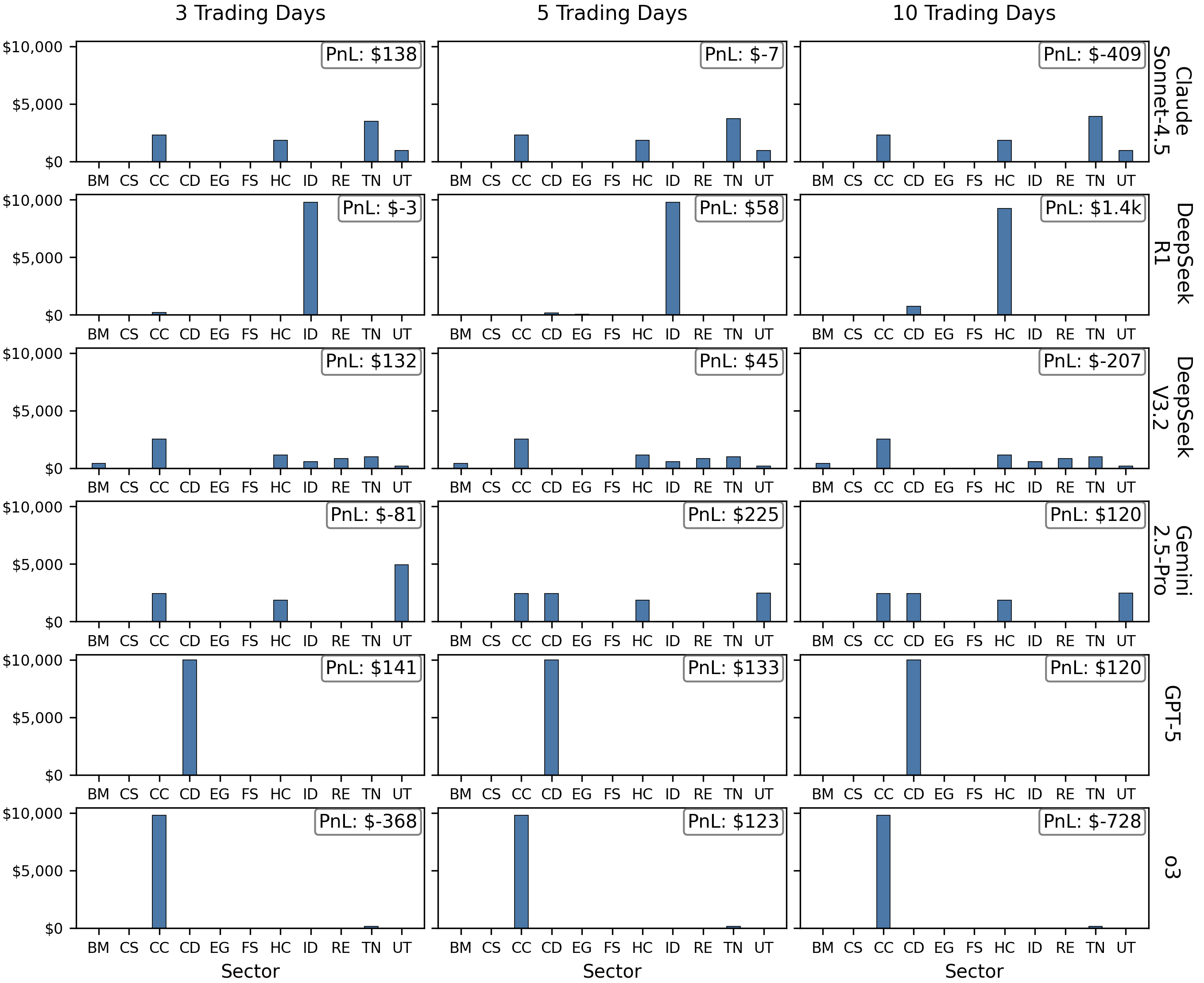}
    \caption{The distribution of investment strategies. ``PnL'' denotes the profit and loss.}
    \label{fig:return}
\end{figure}

\subsection{Main Results}
\textit{1) LLMs' performance varies across different prediction horizons.}
Table \ref{tab:main_table} shows the stock prediction performance across three horizons of six LLMs. First, DeepSeek-V3.2 achieves the lowest overall relative error in the short-term prediction ($\eta=2.14$), excelling particularly in energy, financial services, and industrials.
In the medium-term horizon, GPT-5 delivers the best overall performance ($\eta=2.53$) and leads in multiple sectors, including consumer cyclical, financial services, real estate, technology, and utilities. Similarly, GPT-5 again obtains the lowest overall relative error ($\eta=4.3$) in long-term prediction, achieving the best performance on seven sectors.
It is worth noting that the average hit rates of short-, medium-, and long-term are 0.6, 0.53, and 0.51, respectively, demonstrating the potential in predicting stock price trends towards substantial profits.

\textit{2) Innovation-driven sectors vs. necessity-based sectors.}
Fig. \ref{fig:sector_fig} presents the sector-specific forecasting results, revealing a clear discrepancy in predictability.
We can observe that those innovation-driven sectors, such as communication services, technology, healthcare, and consumer cyclical, exhibit high uncertainty due to rapid disruption, non-linear growth, and event-specific volatility.
In contrast, stable sectors, such as consumer defensive, energy, and utilities, show relatively consistent accuracy, benefiting from inelastic demand, transparent regulation, and strong macroeconomic correlation. Notably, healthcare maintains persistently high error across horizons, reflecting its unique, event-driven risk profile.
 
\textit{3) Short-term homogeneity, medium-term repricing, and long-term bifurcation.}
As shown in Fig. \ref{fig:pearson_fig}, cross-sector forecast correlations evolve distinctively across prediction horizons. 
Short-term results display widespread positive correlation, largely driven by common market volatility, apart from healthcare.
In the long-term prediction, correlations bifurcate into a positively correlated {\it real-economy} cluster and a negatively correlated {\it growth} cluster, reflecting potential risk‑on/risk‑off dynamics. 
The medium-term prediction window exhibits mixed and unstable patterns, indicating a transitional phase of market repricing between short- and long-term regimes.

\textit{4) Risk-seeking group vs. risk-averse group.}
In Fig. \ref{fig:return}, we visualize the distribution of investment strategies of six LLMs, which 
fall into two distinct styles. 
One group, comprising DeepSeek‑R1, GPT‑5, and o3, concentrates most capital in a single sector, reflecting a high‑conviction, high‑risk approach that yields both the highest returns and largest losses in the experiment.
The remaining three models employ a diversified strategy, spreading investments across multiple industries to mitigate risk and pursue more stable returns, resulting in consistently moderate performance.

\begin{table}[t]
\centering
\caption{Impact of financial indicators, news, and fake news. $T_N$ denotes predictions on $N$ trading days.}
\resizebox{.9\linewidth}{!}{\begin{tabular}{cccc|ccc}
\hline
Raw & Indicator & News & Fake News & $T_3$ & $T_5$ & $T_{10}$ \\ \hline
$\checkmark$ &  &  &  & 2.349 & {\bf 2.801} & {\bf 4.822} \\
$\checkmark$& $\checkmark$&  &  & 2.275 & \underline{2.818} & \underline{4.931} \\
$\checkmark$&  & $\checkmark$&  & {\bf 2.151} & 2.943 & 5.556 \\
$\checkmark$ &  &  & $\checkmark$ & 2.449 & 3.109 & 5.242 \\
$\checkmark$ & $\checkmark$ & $\checkmark$ &  & \underline{2.206} & 2.895 & 5.268 \\
$\checkmark$ & $\checkmark$ &  & $\checkmark$ & 2.436 & 2.877 & 4.994 \\ 
\hline
    \end{tabular}}
    \label{tab:abl}
\end{table}

\begin{figure}[t]
    \centering
    \includegraphics[width=.5\textwidth]{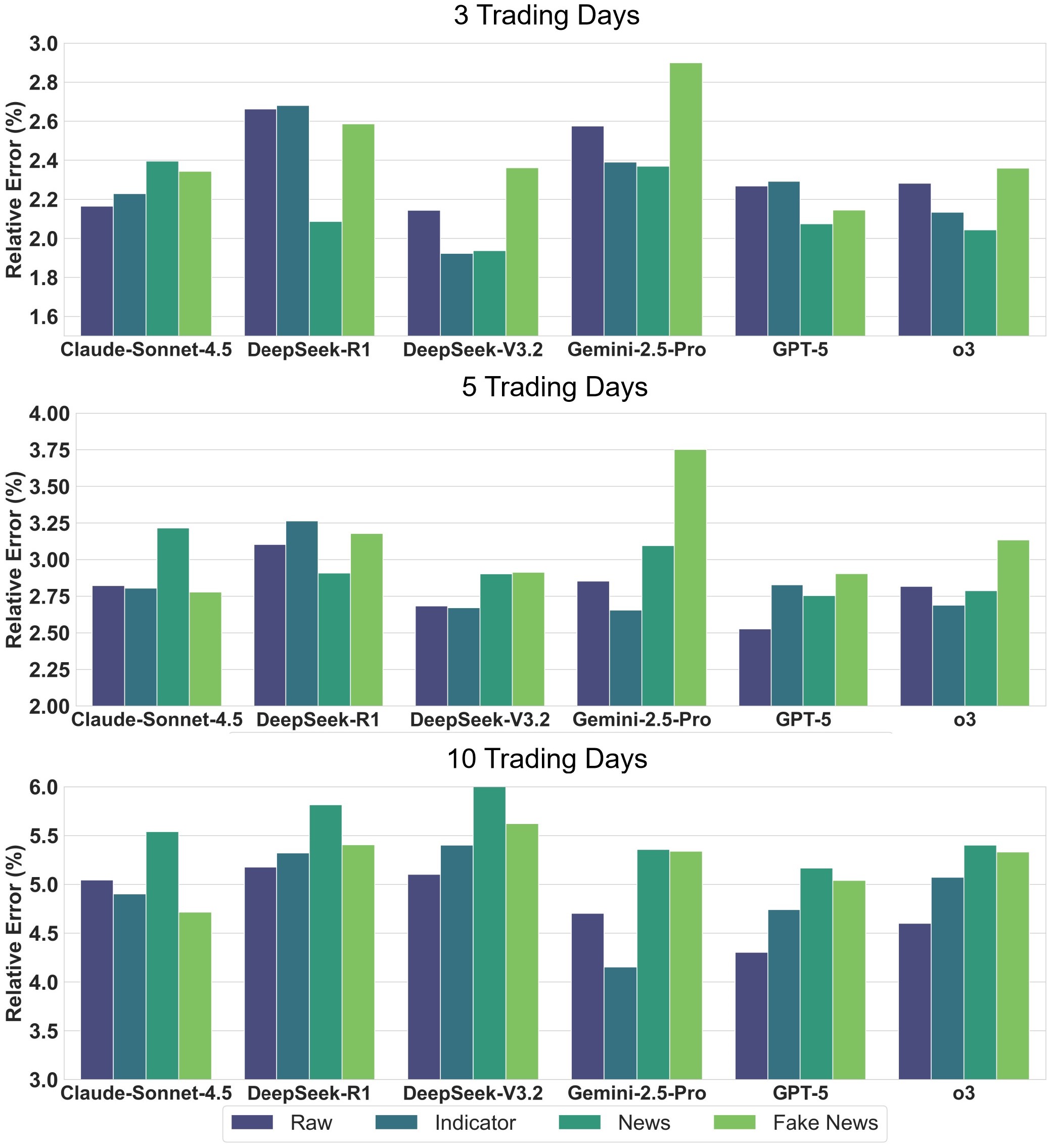}
    \caption{The result of ablation studies across models.}
    \label{fig:ablation}
\end{figure}

\noindent    
\subsection{Ablation Study}
We further investigate the effect of financial indicators, news, and fake news independently and comprehensively on the prediction accuracy, as shown in Table \ref{tab:abl} and Fig. \ref{fig:ablation}.
We find that auxiliary information can significantly affect predictions. 
Both financial indicators and real news enhance short‑term accuracy, with news providing a stronger signal, reducing the relative error by 0.074 and 0.198, respectively.
However, their inclusion harms medium- and long‑term predictions, indicating their effects are transient and confined to near‑term pricing. 
Moreover, fake news severely degrades short‑term performance, yet this negative influence fades over longer horizons and can even turn slightly beneficial, as models gradually learn to identify and discount its exaggerated or fabricated content.
The model-wise comparisons in Fig. \ref{fig:ablation} also demonstrate this observation.

\section{Conclusion}
In this paper, we present PriceSeer, a real-time, production-grade benchmark that challenges large language models to forecast stock price movements. Spanning 110 U.S. equities across 11 sectors, PriceSeer provides 249 chronologically ordered data points per stock, encompassing the opening price, closing price, daily high price, daily low price, and trading volume. We also investigate the impact of internal historical data, external news, and fake news. Experiments on cutting-edge LLMs demonstrate their significant potential in providing accurate stock predictions while pointing out their respective advantages in different sectors and prediction horizons.


\end{document}